\documentclass[%
 aip,
 nofootinbib,
 amsmath,amssymb,
 reprint,%
]{revtex4-1}

\usepackage{graphicx}
\usepackage{dcolumn}
\usepackage{bm}
\usepackage[utf8]{inputenc}
\usepackage[T1]{fontenc}
\usepackage{mathptmx}
\usepackage{etoolbox}
\usepackage{url}
\usepackage{mathtools}
\usepackage{tikz}
\tikzset{>=latex}
\usepackage{color}

\usepackage[utf8]{inputenc}
\usepackage{fullpage}
\usepackage{amsmath,amsfonts,amsthm,amssymb}
\usepackage{url}
\usepackage{booktabs}
\usepackage{bm}
\usepackage{stackrel}
\usepackage{enumitem}

\usepackage[parfill]{parskip}

\newcommand{\Xc}{{\mathcal{X}}}

\makeatletter
\def\@email#1#2{%
 \endgroup
 \patchcmd{\titleblock@produce}
  {\frontmatter@RRAPformat}
  {\frontmatter@RRAPformat{\produce@RRAP{*#1\href{mailto:#2}{#2}}}\frontmatter@RRAPformat}
  {}{}
}%
\makeatother

\begin{document}

\preprint{AIP/123-QED}

\title{Systematic Analysis of Biomolecular Conformational Ensembles with PENSA} 



\author{Martin V{\"o}gele}
\thanks{Equal contributions} 
\altaffiliation{Present address: Schr\"odinger Inc., 1540 Broadway 24th Floor, New York, NY 10036, United States}
\email[Authors to whom correspondence should be addressed: ]{martin.voegele@schrodinger.com, ron.dror@stanford.edu}
\affiliation{Department of Computer Science, Stanford University}
\affiliation{Department of Molecular and Cellular Physiology, Stanford University School of Medicine}
\affiliation{Department of Structural Biology, Stanford University School of Medicine}
\affiliation{Institute for Computational and Mathematical Engineering, Stanford University}
\author{Neil J. Thomson}
\thanks{Equal contributions}
\affiliation{Department of Computational Biology, University of Dundee}
\author{Sang T. Truong}
\affiliation{Department of Computer Science, Stanford University}
\author{Jasper McAvity}
\affiliation{Department of Computer Science, Stanford University}
\author{Ulrich Zachariae}
\affiliation{Department of Computational Biology, University of Dundee}
\affiliation{Department of Biological Chemistry and Drug Discovery, University of Dundee}
\author{Ron O. Dror}
\affiliation{Department of Computer Science, Stanford University}
\affiliation{Department of Molecular and Cellular Physiology, Stanford University School of Medicine}
\affiliation{Department of Structural Biology, Stanford University School of Medicine} 
\affiliation{Institute for Computational and Mathematical Engineering, Stanford University}

\date{\today}

\begin{abstract}
    Atomic-level simulations are widely used to study biomolecules and their dynamics. A common goal in such studies is to compare simulations of a molecular system under several conditions --- for example, with various mutations or bound ligands --- in order to identify differences between the molecular conformations adopted under these conditions. However, the large amount of data produced by simulations of ever larger and more complex systems often renders it difficult to identify the structural features that are relevant for a particular biochemical phenomenon. 
    We present a flexible software package named PENSA that enables a comprehensive and thorough investigation into biomolecular conformational ensembles. It provides featurizations and feature transformations that allow for a complete representation of biomolecules like proteins and nucleic acids, including water and ion binding sites, thus avoiding bias that would come with manual feature selection. 
    PENSA implements methods to systematically compare the distributions of molecular features across ensembles to find the significant differences between them and identify regions of interest.
    It also includes a novel approach to quantify the state-specific information between two regions of a biomolecule, which allows, e.g., tracing information flow to identify allosteric pathways.
    PENSA also comes with convenient tools for loading data and visualizing results, making them quick to process and easy to interpret.
    PENSA is an open-source Python library maintained at \url{https://github.com/drorlab/pensa} along with an example workflow and a tutorial. 
    We demonstrate its usefulness in real-world examples by showing how it helps to determine molecular mechanisms efficiently.
\end{abstract}

\maketitle


\section{Introduction}

Molecules exist not as static structures, but in a range of conformations that fluctuate about energetic equilibria and can be described as a thermodynamic ensemble. 
In recent years, molecular dynamics (MD) simulations have become one of the standard methods in molecular biology,\cite{Hollingsworth2018} providing detailed insights into a molecule's conformational ensemble, complementing static experimental structures that represent only the most probable conformation.\cite{Knoverek2018, Suomivuori2020}
A common and important problem in molecular biophysics is to compare the behavior of a macromolecule under two or more conditions and identify the resulting conformational differences. Typical analyses include investigating the effects of small-molecule ligand binding,\cite{Provasi2011, McCorvy2018} protein mutations,\cite{Cordero-Morales2007} protonation state changes,\cite{Liu2015}, or ion binding events,\cite{Thomson2024} usually by simulating the macromolecule with and without a particular ligand, mutation, or proton.
To derive causal interrelations, the resulting differences are usually identified by visual inspection.\cite{humphrey1996a, DeLano2002} 
However, in systems where the subtleties of a small shift in the populations of minority conformational states or the motion of single atoms might have crucial functional outcomes, analysis is complicated by the large dimensionality of most systems (i.e., the large number of coordinates required to represent the positions of all the atoms). \cite{Hollingsworth2018} 
Even when the dimensionality is distilled to a subset of features, such as torsion angles, simulations contain billions of frames, and functionally relevant differences can extend to regions far from a small-molecule binding site or mutation site.\cite{Provasi2011, Dror2011, Zhou2019, Dror2013, Bowman2015, Hollingsworth2018} 
The necessary analysis is the bottleneck of many biomolecular simulation projects, as it can take weeks of dedicated work if performed by eye and by one-off scripts, and a focus on preconceived candidate mechanisms can lead to missing unexpected effects.

In light of these hurdles, the strong interest in ensemble analyses over the past two decades has led to development of ensemble databases with inbuilt analysis tools\cite{Zivanovic2020} and the availability of more powerful, systematic and quantitative approaches, including: single-score similarity measures between two ensembles,\cite{Brueschweiler2003, Lindorf-Larsen2009} implemented in libraries such as Encore;\cite{Tiberti2015} generative deep learning methods;\cite{Noe2019} Markov model analyses; \cite{Husic2018, Nuske2017} and neural-network based analyses.\cite{Fraccalvieri2011, Ward2021} For example, DiffNets\cite{Ward2021} successfully identified mutation sites that affect the signalling profile of the oxytocin receptor.\cite{Malik2021}
However, the available methods are generally computationally costly, difficult to apply, and/or not easily interpretable. They also frequently require fine-tuning for a particular molecular system.
For proteins in particular, various efficient computational tools have been developed for specific tasks: ProDy\cite{Bakan2011,Zhang2021} specializes on the principal component analysis and normal mode analysis of proteins. It works in cartesian coordinates and thus does not allow for custom featurization. ConsEnsX\cite{Dudola2017} is a web server specifically designed to identify the sub-ensemble best reflecting the data from NMR experiments. EnGens,\cite{Conev2023} a method for the generation of protein ensembles, also includes a wide array of dimensionality reduction functionality to analyze those. Due to their specific purpose, none of these tools are easily transferable to other biomolecules though. The most flexible library for protein dynamics analysis so far has been PyEmma\cite{Scherer2015} which includes functionality to construct Markov models of the protein’s dynamics but it is not actively maintained anymore.
The shortage of standardized yet flexible analysis tools, particularly for comparison of of simulations under multiple conditions, poses an obstacle in many research avenues.

\begin{figure}[hbt!]
    \centering
    \includegraphics[width=\columnwidth]{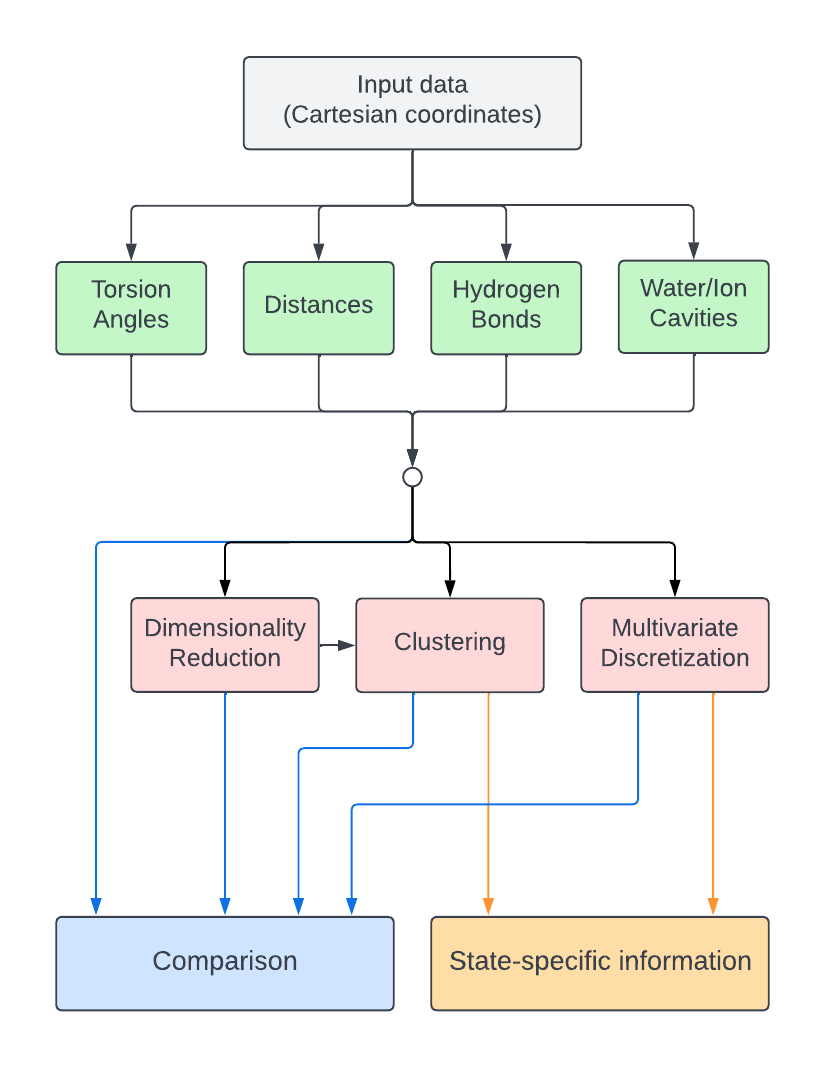}
    \caption{PENSA allows a variety of analysis workflows. Green boxes show primary features (section~\ref{sec:featurization}) and red boxes show possible feature transformations (section~\ref{sec:feature-transformations}). The blue box represents comparison using either the Jensen–Shannon distance (JSD) or the Kolmogorov Smirnov Statistic (KSS), and the yellow box represents the quantification of mutual information using State-Specific Information (SSI) or State-Specific Co-information (CoSSI), which are discussed in sections~\ref{sec:ensemble-comparison} and ~\ref{sec:mutual_inf}, respectively. PENSA's modular nature allows for flexibility in the path from input data to analysis metric.
    }
    \label{fig:workflow}
\end{figure}

We present the modular software library PENSA\cite{pensa} (short for Python ENSemble Analysis) that enables the flexible implementation of systematic and quantitative yet easily interpretable workflows for exploratory analysis of biomolecular conformational ensembles (Fig.~\ref{fig:workflow}). It contains user-friendly tools to preprocess simulation data, to apply various analysis methods across simulation conditions, and to visualize the results. PENSA is an open-source Python library maintained at \url{https://github.com/drorlab/pensa} along with an example workflow and a tutorial.

\begin{figure*}[hbt]
    \centering
    \includegraphics[width=0.8\textwidth]{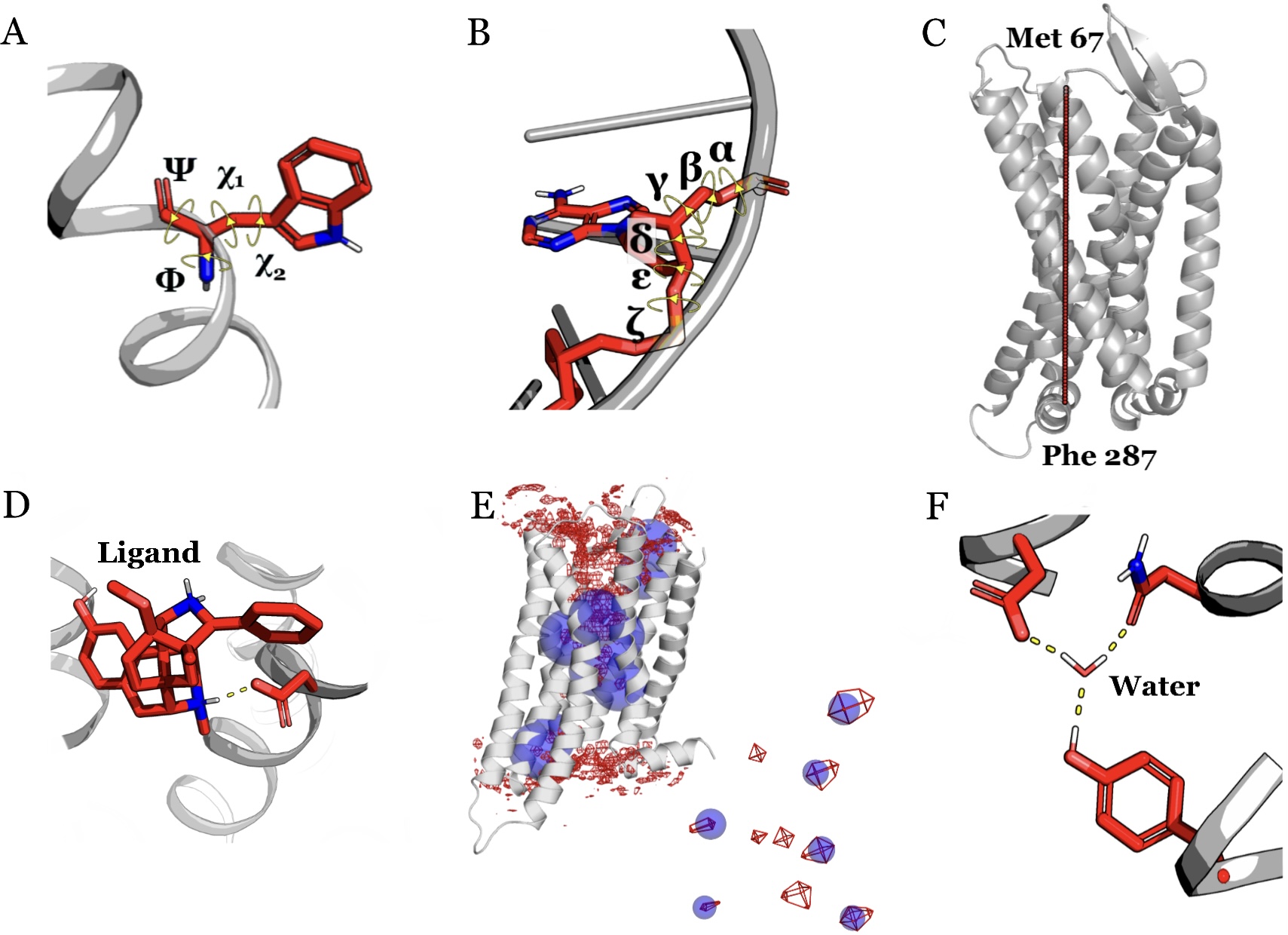}    
    \caption{Examples for biomolecular features implemented in PENSA.
    \emph{A}: Amino acid torsions. $\Phi$ and $\Psi$ represent amino acid backbone torsion angles, while $\chi_i$ represent amino acid side chain torsion angles, as depicted on Tryptophan.
    \emph{B}: Representative angles for DNA and RNA. Six backbone-torsions sufficiently represent the torsional modes of DNA and RNA. 
    \emph{C}: Distances between backbone C$\alpha$ atoms of residue pairs. The illustrated distance captures divergences between ensembles in an example study seen in section \ref{sec:example-receptor}.
    \emph{D}: General hydrogen bonds between two atom groups, such as ligand and protein as depicted in this example.
    \emph{E}: Description of top 10 most probable water sites featurized from water density surrounding the protein accompanied with close up of featurization.
    \emph{F}: Hydrogen bonds of a water molecule within its binding site. 
    All structures are visualized using PyMol\cite{PyMOL} and density features using Biotite.\cite{Kunzmann2018}}
    \label{fig:features}
\end{figure*}

PENSA represents a molecular system via features that allow for its complete representation (Fig.~\ref{fig:features}).
A typical PENSA workflow first determines the same features for all ensembles, currently including: torsion angles of amino acids or nucleic acids, arbitrary interatomic distances, and a novel featurization method for water and ion binding sites. 
The library provides the necessary functions to calculate these features while its modular structure allows for the addition of custom features.
By taking into account all features, PENSA attenuates the bias that would come with manual pre-selection.
To further reduce the system's complexity, methods for dimensionality reduction and clustering can be applied across the joint ensemble combining all conditions, and multiple primary features can be combined to one via multivariate discretization.
The outputs of the feature readers and dimensionality reduction are designed to be compatible with the popular PyEmma library.\cite{Scherer2015}
The features form the basis for the subsequent quantitative analysis that provides comprehensive insight into the ensembles (Fig.~\ref{fig:workflow}).

PENSA's analysis methods focus on quantitatively exploring multiple conformational ensembles and discovering interrelations within and between them. 
To locate the most relevant differences between two ensembles, PENSA includes a direct comparison of every feature's distribution between ensembles via Jensen-Shannon distance (JSD) and the Kolmogorov-Smirnov statistic (KSS). 
Furthermore, a mutual information analysis based on state-specific information (SSI),\cite{Thomson2024} is included to provide a measure of the information that features signal about the ensembles' conditions or the transitions between them. 
An extension of SSI to three variables, CoSSI, enables the tracing of information flow between two regions through the rest of the system.
While mostly designed with the comparison of two simulation conditions in mind, PENSA's analysis methods can be further expanded to more than two ensembles.
For an accessible interpretation of the analysis, PENSA provides various visualization options  that conveniently transfer results to plots or heat maps, or project results onto three-dimensional reference structures.
With these methods, PENSA makes it easy to perform quantitative and systematic exploratory analyses across multiple conditions for a variety of biomolecular systems.

Here, we discuss the functionality included in PENSA and demonstrate its usefulness on three real-world applications: We show how to describe the influence of local frustration on loop opening during the catalytic cycle of an oxidoreductase,\cite{Stelzl2020} how to quantify the influence of force-field parameter changes on simulations of nucleic acids,\cite{Cruz-Leon2021} and how to identify the effect of ionization on receptor proteins.
The examples reproduce existing results, confirming the reliability of our approach, but also demonstrate additional new discoveries made possible by systematic comparison of molecular simulations.


\section{PENSA Functionality}
\label{sec:methods}

\subsection{Featurization}
\label{sec:featurization}
In general, the 3D structure of a molecule is represented with Cartesian coordinates, but in molecular dynamics, this representation is mostly redundant due to bonded and non-bonded atomic constraints.
Other representations summarise the motion of multiple atomic coordinates in a succinct and instructive manner. In the following, we use the term \textit{feature} for every such numerical attribute calculated from the coordinates of a structure. 
PENSA represents each ensemble (e.g., simulation trajectory) by the same set of $F$ features $\{x_1, ..., x_F \}$.
Here we discuss the types of features we consider most suited for a systematic analysis of molecular structure, currently including torsion angles, interatomic distances, water and ion binding sites (for an overview, see Fig.~\ref{fig:features}). They are implemented in PENSA using MDAnalysis\cite{michaud-agrawal2011a, gowers2016a} and can be replaced or extended by other descriptions. PENSA is modular and new featurization functions can be added easily.

\emph{Representing Molecules by Torsion Angles:}
Because the energy minima of molecular structures are largely determined by stereochemistry, the 3D conformation of most biological macromolecules can be approximately described using a set of torsion angles. 
For proteins, we use torsions around chemical bonds to describe the effective degrees of freedom in the protein backbone and the side chains (Figure~\ref{fig:features}). The three rotatable bonds in the protein backbone are characterized by angles $\phi$, $\psi$, and $\omega$. As $\omega$ is almost always in trans configuration, we only use $\phi$ and $\psi$. In addition, the amino acid side chains have up to five rotatable bonds. This results in a number of variables that grows linearly with the number of residues, describing the full protein structure while avoiding the redundancy that comes with using Cartesian coodinates (Fig.~\ref{fig:features}A). 
The analogous description of nucleic acids via torsion angles requires the definition of pseudo-torsions.\cite{Keating2011} The 3D structure of DNA or RNA can be described via the six main chain torsion angles ($\alpha$, $\beta$, $\gamma$, $\delta$, $\epsilon$, $\zeta$) around the covalent bonds and the angle $\chi$ about the glycosidic bond and the sugar pucker (Fig.~\ref{fig:features}B). An alternative description --- also implemented in PENSA --- uses two pseudo-torsions, $\eta$ and $\theta$, that are defined around imaginary lines connecting more distant atoms.\cite{Keating2011}

\emph{Representing Molecules by Distance:}
The set of all distances between all atoms provides a complete description of the structure of a molecule, independent of the coordinate system. However, taking into account all $N$ atoms is extremely inefficient as the number of distances required grows $\propto N^2$ while the number of degrees of freedom grows $\propto N$. 
Reducing the set of distances to a few relevant ones can make this approach much more efficient. For example, the overall structure of a protein is usually described by the distances between all C$\alpha$ atoms. A further reduction in redundancy can be achieved by using system-specific knowledge, obtaining a subset that is smaller but still representative for the overall dynamics of the system. Common examples for this are hydrogen bond lengths to describe bonding patterns or base-pair distances in DNA.
A combined analysis of distances and torsions can be beneficial as some effects can be easier to spot in distances and others in torsions. 

\emph{Representing Hydrogen Bonds:}
Because hydrogen bonds are a fundamental property of protein interactions, PENSA also includes the option to featurize all hydrogen bonds between two distinct groups of atoms, for example, a ligand and a protein. 
All hydrogen bonds are recognized between donor and acceptor atoms using a hydrogen bond cutoff of 3.0~{\AA} and a donor-hydrogen-acceptor angle cutoff of 150$^{\circ}$, using the hydrogen bond module of MDAnalysis.\cite{michaud-agrawal2011a,Smith2019}
Since the angle criterion can be quite expensive to calculate across large input ensembles, PENSA also provides functions to quickly scan for H-bonds based on an older, more lenient criterion using only a distance cutoff of 3.5~\AA.\cite{Kajander2000}
A binary timeseries distribution is generated for every hydrogen bond pair, representing when the bond is present (1) or absent (0), and indexed by the atom names, residue names and sequence numbers. 

\emph{Representing Water Binding Sites:}
We have implemented a new method to represent the presence and orientation of internal water molecules, which are crucial to the structure and function of many biomolecules. Typically, internal water molecules are defined by the protein sites they bind to, or by an atomic density that averages the motion of the waters.\cite{Venkatakrishnan2019, Yuan2014, Pardo2007, Levy2004}
PENSA adds to this with a dynamic, orientation-based representation of water in binding sites within a biomolecule. 
Because water binding sites are often accessible to water molecules in a freely diffusing bulk solvent, water molecules may interchange in a water binding site while the site itself maintains its function. 
Therefore, featurization of a certain water \emph{site} within the biomolecule is more important than the individual water molecules that occupy it.  
The polar nature of water molecules enables a water binding site to function as a polarizable interface, mediating hydrogen bond networks between amino acids that would otherwise be impossible to form.
Any site that can accommodate a water molecule can additionally be unoccupied, whereby its occupation status (occupied vs. empty) can act as a further feature. 
To account for both of these effects, PENSA featurizes a water binding site via its occupation and, when occupied, the orientation of the water molecule. 
First, we locate the water binding sites by finding all local maxima within a 3D density grid. The grid is obtained by aligning the assembled trajectories such that the density refers to the combined ensemble and can be used to compare identical sites across the ensembles. Each local maximum marks the center of a water binding site, defined as a sphere of radius 3.5~\AA, based on the range of typical hydrogen bonding interaction distances.\cite{Kajander2000}
For each simulation frame, the feature value is defined by the orientation of the water molecule (if occupied), or a value representing its unoccupied state.  
The water molecule's orientation is represented by the angular components of its dipole moment in spherical coordinates. This representation is appropriate when rotation, translation, and periodicity effects of the simulation system during the simulation are excluded from the water pocket featurization during preprocessing.

\emph{Representing Hydrogen Bonds in Water Binding Sites:}
Water molecules at specific binding sites mediate local hydrogen bonding interactions between surrounding protein residues.\cite{Venkatakrishnan2019} 
The orientation of the water molecule relative to the protein may be affected by a change in simulation condition, which could have important functional outcomes. 
PENSA provides the option to featurize all hydrogen bonds wihtin the water site to account for these water-protein hydrogen bonds formed in the binding site.
This can be used, for example, to calculate hydrogen bond frequencies under different simulation conditions, or to identify how those precise moments in the simulation timeline where bonds may break or form might temporally couple to other events. 

\emph{Representing Ion Binding Sites:}
PENSA provides a similar approach to featurize ion binding sites, which play important roles in the structure and function of many biomolecular systems --- from ion channels to catalytic enzymes \cite{Zarzycka2019, Andreini2008}. 
Simulation trajectories must be preprocessed in the same manner as for water binding sites, ensuring that the ion density grid is not affected by any system rotation, translation or switches between periodic boundaries. 
Ion binding sites are then represented as a binary feature describing the presence or absence of an ion in the binding site, or as a discrete feature describing the ion indices that occupy it, for instance in order to keep track of functionally relevant ion binding and unbinding events.
This representation allows for an investigation of the movement of specific ions in and out of an array of ion binding sites, and has already been employed to identify co-operative knock-on permeation of ions bound to different sites of cation channels.\cite{Ives2023}

\subsection{Feature Transformations}
\label{sec:feature-transformations}

\begin{figure}
    \centering
    \includegraphics[width=1.0\columnwidth]{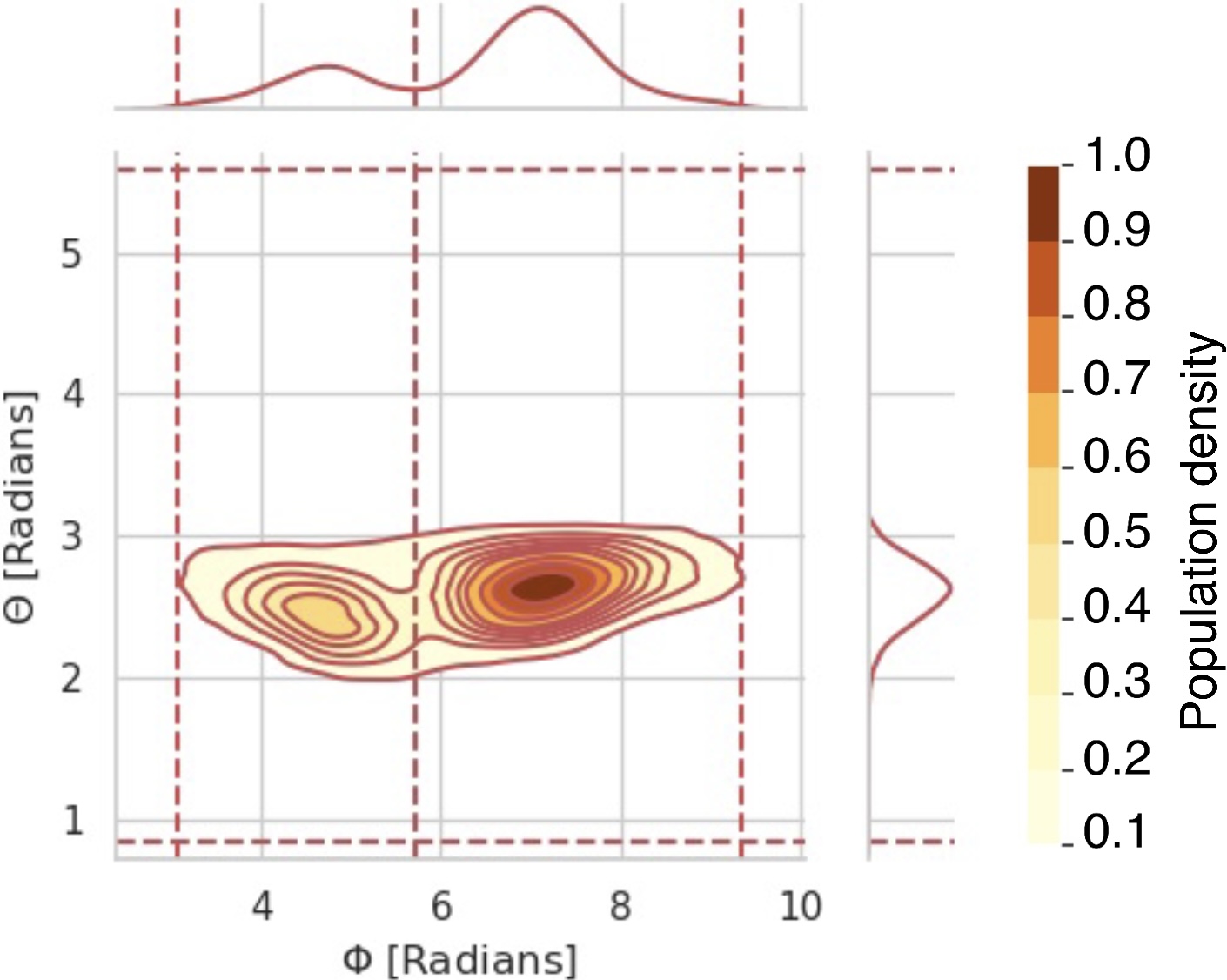}
    \caption{Example for multivariate discretization. Discrete macrostates of a water binding site generated by discretization of individual pairs of spherical coordinates $(\phi, \theta)$ of the polar vector of each water molecule within the water binding site. Red dashed lines depict the state boundaries and include the periodic boundaries as the first and last state boundaries.}
    \label{fig:multivariate-discretization}
\end{figure}

\emph{Multivariate Discretization:}
PENSA offers an automatic discretization of one or multiple combined features into distinct states.
For example, amino acid torsions typically oscillate about local minima in populations known as rotamers, often sampling different rotamers in an MD simulation.\cite{Dunbrack2002, Dunbrack1993, Scouras2011}
In the Dunbrack library,\cite{Dunbrack1993} each local rotamer can be accurately represented by a Gaussian distribution around the minimum energy conformation.
Similarly, Gaussians can represent the oscillation of a water molecule's polarization (orientation of its dipole) within a water binding site, while occupancy changes can be conveniently considered as two discrete populations with zero oscillation.
The discrete states of these distributions are generated by applying a multi-modal Gaussian fit. The fit parameters are obtained with a non-linear least squares fit of up to ten Gaussians to each feature distribution using the SciPy library.\cite{2020SciPy-NMeth}
For our purposes, we found this method to be computationally more efficient than alternative methods such as Gaussian Mixture Model. 
The state limits for each distribution are then defined by the Gaussian intersects.
Distributions which have a cyclic periodicity may oscillate about a periodic boundary but show up as two states on different ends of the distribution.
To account for this boundary effect, all periodic distributions can be linearly shifted.
In general, state limits can be defined for any kind of distribution, even non-Gaussian, using suitable clustering algorithms. 
For these cases, PENSA allows the manual input of arbitrary state boundaries.
In instances where the dynamics of an amino acid are described by a combined view of all side-chain angles, e.g., the five side-chain torsions of arginine, an arbitrary number of \textit{N} dimensions can be combined into one joint feature.
The states of an \textit{N}-dimensional feature are then the combination of all discretized microstates, defining a grid in the combined feature space, as shown in Fig.~\ref{fig:multivariate-discretization}. 

\emph{Dimensionality Reduction:}
Comparisons of multiple complex systems can be made more tractable by representing these systems in terms of low-dimensional, higher-order descriptions that summarize a large number of primary features in a small number of quantities. 
The most popular algorithm for such dimensionality reduction is Principal Component Analysis (PCA)~\cite{pca}. Time series, such as those from molecular dynamics simulations, can also be projected via Time-Lagged Independent Component Analysis (TICA)~\cite{tica}.
Classically, dimensionality reduction has been applied on single ensembles. But to detect patterns across ensembles and differences between them, we have to define the same representation for all investigated ensembles. PENSA users can perform dimensionality reduction on the combined data of all ensembles included in the analysis and then compare them along the resulting reduced dimensions.

\emph{Clustering:}
The coordinate space defined by the dimensionality reduction methods discussed above can be clustered into discrete states, again using data from all ensembles.
Clustering the structures from all ensembles in the resulting lower-dimensional space provides discrete states. PENSA implements k-means clustering~\cite{Macqueen67somemethods} and regular-space clustering~\cite{doi:10.1063/1.3565032}, two popular algorithms for this task. Users can calculate populations of the resulting discrete states in each ensemble and compare them.

\subsection{Feature-by-Feature Comparison}
\label{sec:ensemble-comparison}
The local extent of deviations between two ensembles can be quantified by comparing each feature's probability distribution in one ensemble to its distribution in the other ensemble. 
For a feature $x_f$ in a simulation trajectory of length $T$, the $T$ samples of $x_f$ give an empirical estimation of the distribution $p(x_f)$, which describes the behaviour of $x_f$ in that ensemble.
We want to compare the distribution $p_i(x_f)$ in ensemble $i$ to the corresponding distribution $p_j(x_f)$ in ensemble $j$. These distributions may have complex functional forms. Thus, comparing their summary statistics (mean and standard deviation) might not be sufficient because they cannot capture more subtle differences in the distributions, e.g., the split-up of one state (unimodal distribution) into two states (bimodal). Instead, PENSA provides comparison measures that are designed to capture differences in probability distributions, namely the Jensen-Shannon distance (JSD) and the Kolmogorov-Smirnov statistic (KSS).

\emph{Jensen-Shannon Distance:}
Two distributions can be compared using the Jensen-Shannon distance $D_{JS}$, a symmetrized and numerically more stable version of the Kullback-Leibler divergence $D_{KL}$. For two distributions over a feature $x_f$ from ensemble $i$ and ensemble $j$, $D_{JS}$ is defined below
\begin{eqnarray}
    D_\mathrm{JS}[p_i\|p_j] &=& \sqrt{\frac{1}{2} \left( D_{KL}[p_i\| \bar{p}] + D_{KL}[p_j\| \bar{p}] \right) } \\
    \mathrm{with}\; \bar{p} &=& \frac{p_i + p_j}{2} \nonumber
\end{eqnarray}
with $D_{KL}$ the Kullback-Leibler divergence. For numerical reasons, we always use its discrete version:
\begin{align}
    D_{KL}[ p_i(x_f) || p_j(x_f) ] &= \int p_j(x_f) \log\frac{p_i(x_f)}{p_j(x_f)} dx_f \\
    &\approx \sum_{x_f\in\mathcal{X}_f} p_j(x_f) \log\frac{p_i(x_f)}{p_j(x_f)} \nonumber 
\end{align}
where $\Xc_f$ is the set of possible states. In the case of continuous features, these states are bins along the feature coordinate, obtained by evenly dividing the range of the joint distribution. Note that $D_{KL}$ is not symmetric, i.e., $D_\mathrm{KL}[p_i \parallel p_j] \neq D_\mathrm{KL}[p_j\parallel p_i]$ but $D_{JS}$ is.
The use of JSD as a comparison metric has been discussed in more detail in previous work\cite{Lindorf-Larsen2009} where the comparison was performed on entire ensembles instead of individual features.
Most importantly, in contrast to the unbounded --- and in practice often divergent --- KL divergence, the Jensen-Shannon distance ranges from 0 to 1 where 0 is obtained for identical distributions and 1 for a pair of completely different distributions. 

\emph{Kolmogorov-Smirnov Statistic:} 
Alternatively, we can quantify the deviations between two distributions of continuous features without the need to define a binning parameter by using the Kolmogorov–Smirnov statistic $D_{KS}$. It is defined for a feature $x_f$ as
\begin{equation}
    D_\mathrm{KS}(p_i\| p_j) = \sup_{x_f} \| F_{p_i}(x_f) - F_{p_j}(x_f)\| \label{eq:kss}
\end{equation}
with $F_{p_i}(x_f)$ and $F_{p_j}(x_f)$ the empirical distribution functions of $p_i(x_f)$ and $p_f(x_f)$, respectively. The empirical distribution functions are directly obtained from the calculated features and require no sorting of the data into arbitrary bins. The results of $D_\mathrm{JS}$ and $D_\mathrm{KS}$ for the same comparison ideally are very similar which can serve as an important sanity check.

\emph{Overall Ensemble Similarity:}
The overall similarity of two ensembles over all features in a metrics can be quantified by aggregating similarity metrics of all features $x_f$. For example, an average Kolmogorov-Smirnov statistic $\bar{D}_{KS}$ of two ensembles $i$ and $j$ can be computed as:
\begin{equation}
    \bar{D}_{KS} = \frac{1}{F} \sum_{f=1}^F D_{KS} [p_i(x_f) || p_j(x_f)] \label{eq:aggregate}
\end{equation}
and $\bar{D}_{JS}$ analogously.
Aggregation functions other than the average, including the maximum and the minimum, are also implemented. Similar to other metrics that quantify the similarity of two ensembles in a single score,\cite{Brueschweiler2003, Lindorf-Larsen2009, Tiberti2015} these aggregated metrics are particularly helpful when we need to evaluate the output of a new method to a reference ensemble, for example comparing a simulation or a generative machine learning model to a ground truth from an experiment or a more precise level of modeling.

\subsection{Mutual Information Analysis}
\label{sec:mutual_inf}

\begin{figure}
    \centering
    \includegraphics[width=0.975\columnwidth]{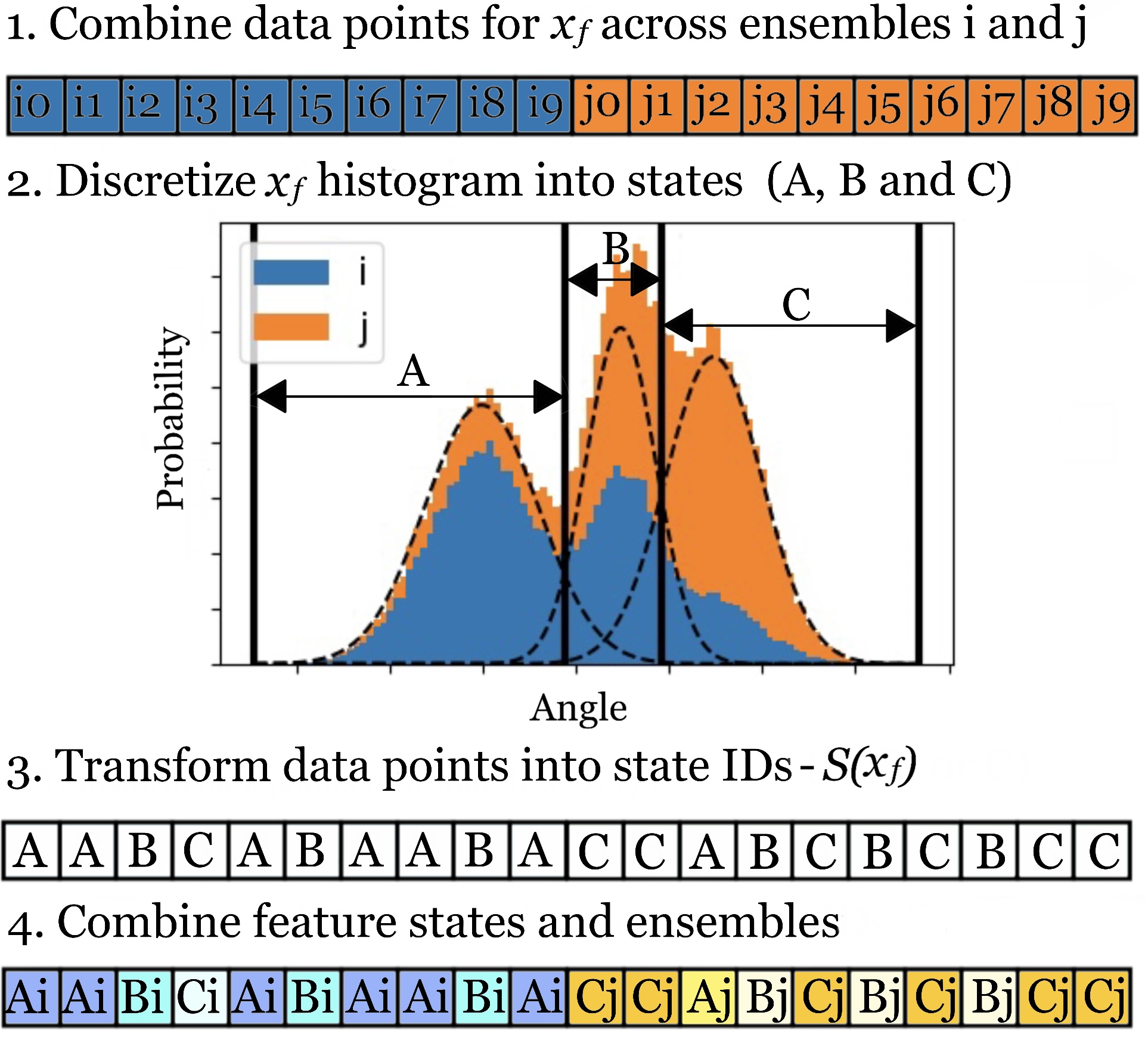}
    \caption{Schematic illustration of the calculation process for the State-Specific Information of feature $x_f$, $I_{SSI}(x_f)$. The combined data points for feature $x_f$ are discretized into states defined by the intersects of a Gaussian fit to the data histogram; blue and orange represent ensembles i and j, respectively. Each data point is transformed into the corresponding discrete state it lies within ($S(x_f)$) and combined with the ensembles - $i$ or $j$. Colors are added to the data points on step 4 to highlight the unique substates that arise from the combination of feature states and ensembles.}
    \label{fig:ssi_workflow}
\end{figure}

A mutual information analysis can be employed to measure how much the specific value of one feature is coupled to the specific value of another.\cite{Shannon1948}
Applying this approach to two conformational ensembles $i$ and $j$, one can identify if the specific values of a feature are dependent on the system ensemble ($i$ \textit{or} $j$), and \textit{vice versa}, how much the value of a feature reveals about whether it stems from ensemble $i$ \textit{or} $j$.
The ensemble identifiers $i$ and $j$ are then equivalent to the values of an additional feature within one (joint) ensemble. 
PENSA focuses on mutual information shared between a feature's conformational states (e.g., the multivariate states described in section \ref{sec:feature-transformations}) and the ensembles $i$ and $j$.
To quantify this, we calculate the State-Specific Information (SSI, Fig.~\ref{fig:ssi_workflow}), a linear, discrete-state adaptation of mutual information that has originally been developed for amino acid torsions that act as molecular switches\cite{Thomson2024} but here is generalized to any feature with a distribution that can be represented as discrete states.
The discretization of protein features is a transformation that maps each time-series value to a discrete state identifier for that time-series distribution. 
An arbitrary number of ensembles could be incorporated into the SSI calculation, simultaneously measuring the mutual information between all conditions, but it is currently implemented for two.
Similarly, SSI can operate on a single ensemble that is partitioned into two sub-ensembles, e.g., along a state boundary.

\emph{State-Specific Information (SSI):}
The SSI measure $I_\mathrm{SSI}(x_f)$ quantifies the degree to which conformational state transitions of feature $x_f$ signal information about the ensembles $i$ and $j$ or the transitions between them. It is defined as
\begin{equation}
    I_\mathrm{SSI}(x_f) =\sum\limits_{\substack{s \in S(x_{f}) \\ e \in [i, j]}} p(s, e) \log\frac{p(s, e)}{p(s) p(e)} \label{eq:ssi}
\end{equation}
where $S(x_f)$ is the transformation of the combined data points of $x_f$ across ensembles $i$ and $j$ into data points referring to a feature state-identifier $s$, derived from the discretized probability distribution of $x_f$, as seen in figure \ref{fig:ssi_workflow}. The variable $e$ is a generic representation of the ensemble ID, with possible values $i$ or $j$. 
$p(s, e)$ is the joint distribution of a feature state of $x_f$ and ensemble (obtained from the probabilities of each element in the list shown in Fig.~\ref{fig:ssi_workflow}, step 4.), and $p(s)$ and $p(e)$ are the marginal distributions.
The SSI ranges from 0 bits to 1 bit, where 0 bits represents no shared information and 1 bit represents maximal shared information between the ensemble (transitions) and the features. 

\emph{State-Specific Co-Information (CoSSI):}
To quantify the degree to which two features interact with one another as they signal information about the ensemble they are in ($i$ or $j$), State-Specific Co-Information (CoSSI) is employed. 
This multivariate feature-feature-ensemble metric is the linear, discrete-state adaptation of co-information\cite{Bell2003} that uses Shannon's discrete entropy formulation,\cite{Shannon1948} calculated using 
\begin{equation}
    \begin{aligned}
    &I_\mathrm{CoSSI} (x_1, x_2)  \\
    &=\sum\limits_{\substack{s_1 \in S(x_1) \\ s_2 \in S(x_2)}}
    p(s_1, s_2) \log\frac{p(s_1, s_2)}{p(s_1) p(s_2)} \\
    &-\sum\limits_{\substack{s_1 \in S(x_1) \\ s_2 \in S(x_2) \\ e \in [i, j])}}
    p(s_1,s_2,e) \log\frac{p(s_1,s_2,e)p(e)}{p(s_1,e) p(s_2,e)} 
\end{aligned}
\end{equation}
where the transformation $S(x_f)$ is as previously defined.
$I_{CoSSI}$ can be positive or negative, indicating whether the switch between ensembles increases ($I_{CoSSI} > 0$), decreases ($I_{CoSSI} < 0$), or does not affect ($I_{CoSSI} = 0$) the communication between two features $x_1$ and $x_2$. In the case of small-molecule ligand binding, for instance, positive $I_{CoSSI}$ between features can represent the turning-on of a signal channel by a ligand.

\subsection{Visualization}

PENSA includes convenient functions to visualize all stages of the analysis workflow.
Primary features as well as processed features (like projections onto PCA eigenvectors) can easily be compared individually using histograms or inferred densities, and combinations of two features using heatmaps, with all functionality based on Matplotlib.\cite{Hunter2007}
While featurizing water and ion binding sites, the average position of the molecules are extracted via MDAnalysis.\cite{michaud-agrawal2011a}
The binding sites' centers and the magnitude of the probability maxima are stored in additional atoms, added via Biotite\cite{Kunzmann2018}.
Input structures can be sorted along the values of primary or processed features using MDAnalysis\cite{michaud-agrawal2011a} which is particularly useful for PCA or tICA to see which component of a molecule's motion is associated with which eigenvector.
Analysis metrics for comparison and mutual information that are related to a single residue (e.g., the maximum JSD of all side-chain torsions in an amino acid) can be stored in structure files using MDAnalysis\cite{michaud-agrawal2011a}.
We provide scripts for PyMol\cite{PyMOL} and VMD\cite{humphrey1996a} to visualize them via the color or the width of the cartoon representation. Metrics related to two features (e.g., distances) are visualized in square heatmaps, also implemented via Matplotlib.\cite{Hunter2007}
These visualization options provide a comprehensive overview of complex systems in one --- or very few --- figures.


\section{Example Applications}

\subsection{Understanding effects of a small chemical modification: loop opening in an oxidoreductase}
\label{sec:example-oxidoreductase}

As a first example, we show how the systematic comparison of protein backbone and side-chain torsions provides a comprehensive overview on the differences in the conformational ensembles induced by a small chemical modification. We consider the oxidation of two cysteine thiols to a disulfide bond in the N-terminal domain of the key bacterial oxidoreductase DsbD (nDsbD). DsbD plays an important role in electron transport across the inner cytoplasmic membrane of gram-negative bacteria and this reaction is an important step in its catalytic cycle.\cite{Stelzl2020}

\begin{figure}
    \centering
    \includegraphics[width=0.7\columnwidth]{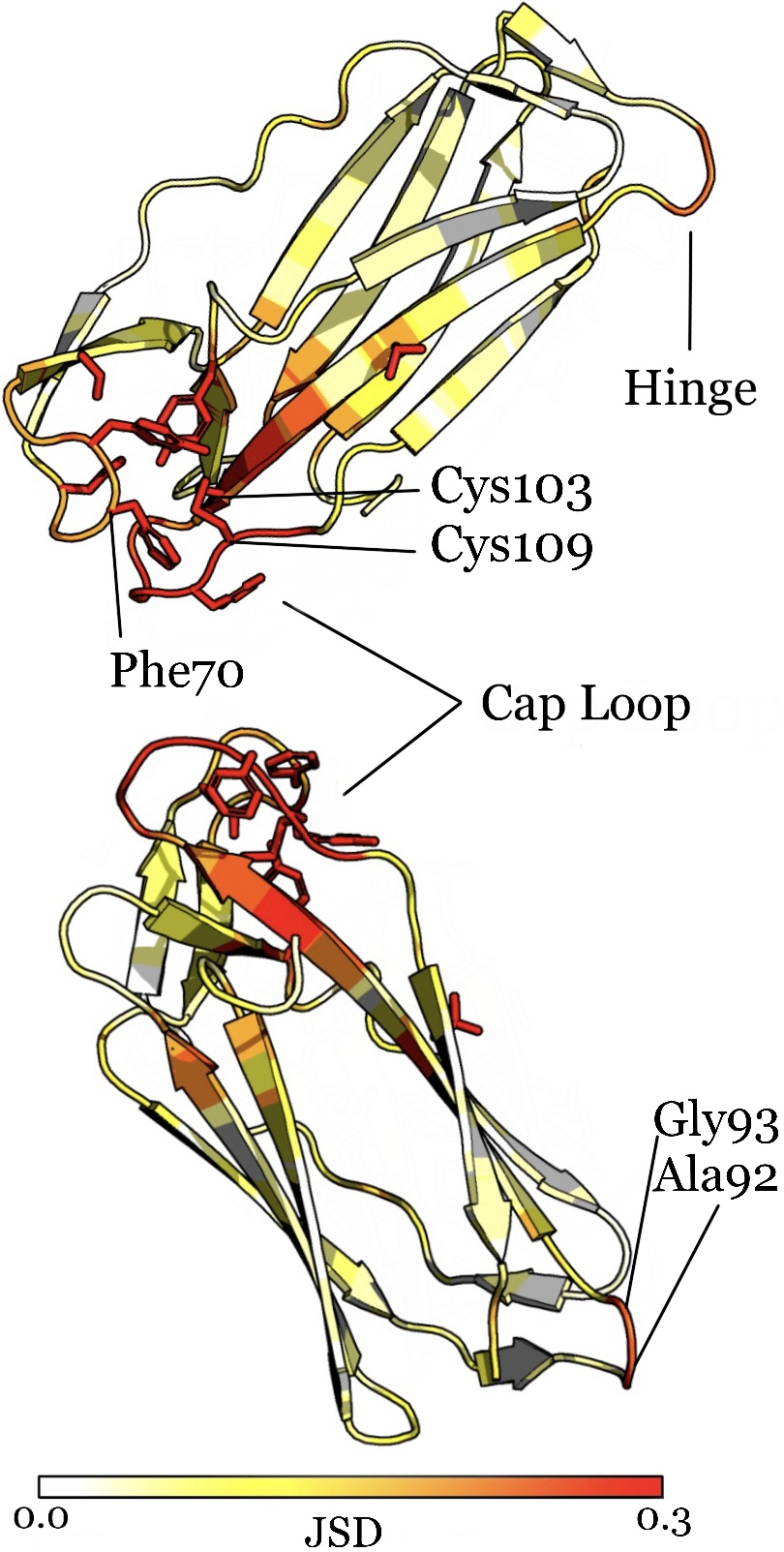}
    \caption{Comparison of torsion angles via the Jensen-Shannon Distance (JSD) reveals where conformational ensembles differ and allows to visualize these effects on a single structure. Analysis of backbone and side-chain torsions highlights regions of nDsbD that are most affected by the oxidation of the two cysteine thiols, Cys103 and Cys109 \cite{Stelzl2020} For each residue, the maximum JSD of the distributions of backbone torsion angles between oxidised and reduced condition is encoded in the color of the cartoon representation from white (0.0) via yellow and orange to red (0.3). Similarly, the maximum JSD for the sidechain torsion angles is encoded in the color of the side chains, which are displayed in stick representation only for residues with JSD values greater than 0.3. 
    Structures are visualized using PyMol.\cite{PyMOL}
    }
    \label{fig:nDsbD-JSD}
\end{figure}

Visualization of the maximum JSD per residue for backbone and sidechain torsions (see Fig.~\ref{fig:nDsbD-JSD}) shows at one glance the opposing residues in the cap loop to be the most affected regions. Unsurprisingly, the residues directly involved in the reaction, Cys103 and Cys109, show the highest values. The effects of this reaction on the surrounding residues -- showing up in our analysis as medium JSD values -- cause a change in the distance of the opposing loop and a corresponding opening of the cap-loop region. In particular, we find residues Phe70 and Tyr71 at the neighboring loop to be strongly affected. Indeed, the authors of the original study identified the distance between residues Phe70 and Cys109 to be the characteristic hallmark of loop opening. 

Besides the influence on the cap loop, we identify a more subtly involved and previously not discussed region at the opposite side of the protein. This region -- mainly the backbones of residues Ala92 and Gly93 -- functions as a hinge for the beta-strand region that slightly tilts when the cap loop is pushed outward during the cap opening. This finding demonstrates how our approach picks up differences between simulations that are otherwise easily missed.

\subsection{Comparing force field parameters: Interactions of Calcium with DNA}
\label{sec:example-dna}
\begin{figure}
    \centering
    \includegraphics[width=0.7\columnwidth]{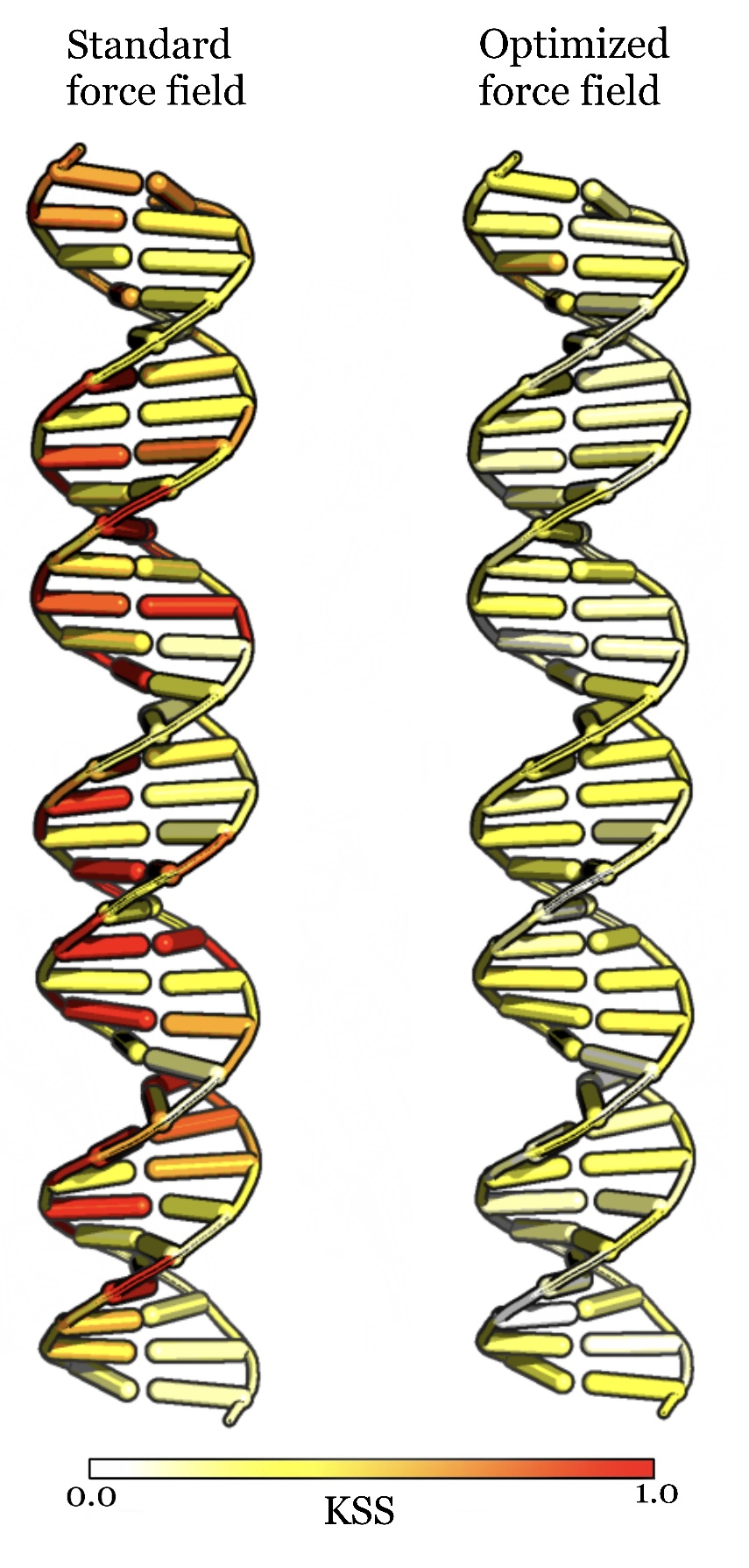}
    \caption{Comparison of torsion angles via the Kolmogorov-Smirnoff Statistic (KSS) reveals where conformational ensembles differ and allows to visualize these effects on a single structure.
    Comparison of DNA backbone torsion angles from MD simulations to an experimental reference ensemble (see main text for details) shows the improvement in the ensemble generated by MD simulations when switching from standard force field parameters (left) to optimized force field parameters (right). Structures are visualized using PyMol.\cite{PyMOL}
    }
    \label{fig:DNA-KSS}
\end{figure}

As a second use case, we show how to quantify the effects of small changes in force field parameters on the overall conformational ensemble (Fig.~\ref{fig:DNA-KSS}). 
We consider the binding of calcium ions (Ca$^{2+}$) to DNA \cite{Cruz-Leon2021}. Metal cations play a crucial role in stabilizing the structure of nucleic acid systems. Their force field parameters are usually determined to reproduce bulk properties like the solvation-free energy and thus often not directly transferable to interactions with biomolecules.\cite{Panteva2015} Furthermore, small changes in the interactions can have significant conformational consequences overall.\cite{Yoo2012} Studying the effect of such small changes on the overall conformation is an important problem in force field optimization.

Comparison of DNA backbone torsions from MD simulations using the standard force field parameters for Ca$^{2+}$-DNA interactions to an experimental reference ensemble reveals a periodic pattern of strong deviations along the entire double strand that are strongly reduced by using optimized force field parameters. 
We generate the reference distribution from the experimental structure by Gaussian sampling of the relevant coordinates using the resolution of the X-ray structure (1.7~\AA, PDB:~477D) as the width of the distribution. We then compare the simulations of each parameter set to this reference (Fig.~\ref{fig:DNA-KSS}) using the Kolmogorov-Smirnov statistic (Eq.~\ref{eq:kss}) because it provides a parameter-free measure for the deviations.
The approximate periodicity of the deviations (Fig.~\ref{fig:DNA-KSS}, left) shows that, using standard parameters, the conformational ensemble as a whole deviates from the reference distribution and has problems reproducing the overall structure of the double strand.
The authors of the original study identified an overestimation of Ca$^{2+}$-DNA interactions as the main cause of such deviations.\cite{Cruz-Leon2021} It allows the Ca$^{2+}$ ions to bridge between the phosphate oxygen atoms of opposite backbone strands which causes the minor groove of the DNA strand to shrink and in turn affects the entire structure. Thus, they rescaled the force field parameters to optimize the Ca$^{2+}$-DNA interactions. This not only improved the local accuracy but the entire conformational ensemble, even though the overall structure was not explicitly optimized for during the rescaling. In our PENSA-based analysis, this improvement is immediately visible by the reduced deviations (Fig.~\ref{fig:DNA-KSS}, right). The overall ensemble KSS (as in Eq.~\ref{eq:aggregate}) is reduced from 0.50 (standard force field) to 0.34 (optimized force field).
This example shows that our workflow quickly identifies whether and where small changes in local interactions propagate to strong deviations in the overall structure of a biomolecule.

\subsection{Tracing information linked to a protonation state: The central aspartic acid in the $\mu$-opioid receptor}
\label{sec:example-receptor}

\begin{figure*}[hbt]
    \centering
    \includegraphics[width=\textwidth]{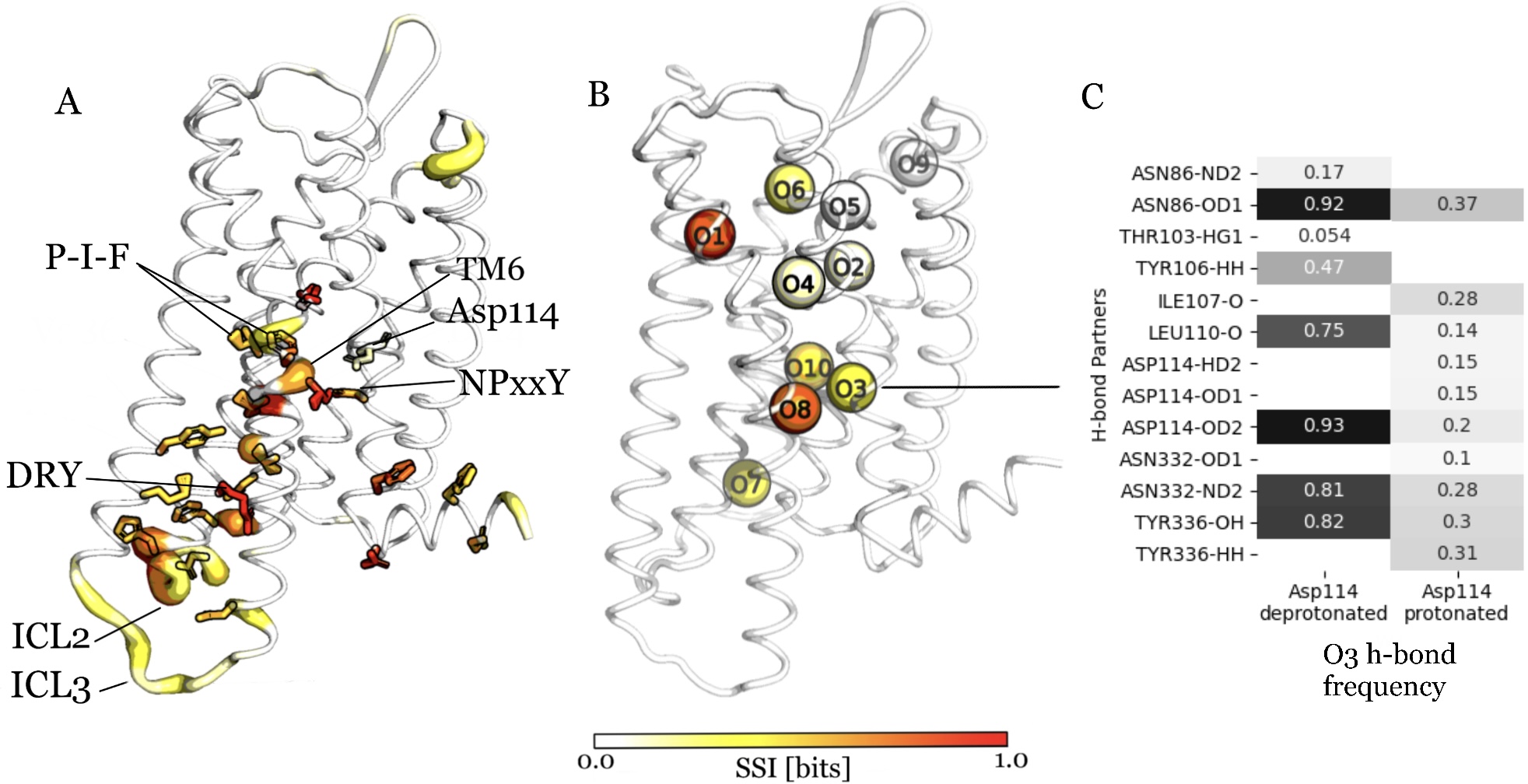}
    \caption{Analysis of backbone and side-chain torsion angles (A, cartoon and stick), and water binding sites (B, spheres), via State-Specific Information (SSI) quantifies the information that each feature signals about the protonation state of Asp114 (D2.50) in the inactive-state murine $\mu$-opioid receptor. 
    SSI highlights that changes in the protonation state of Asp114 couple to rotameric state changes on: the backbone of TM6, proximal to Asp114; the backbone of ICL2 and ICL3; residue side chains on the P-I-F, NPxxY and DRY motifs; and water sites distributed throughout the receptor. The SSI for backbone torsions is encoded in the color and width of the cartoon representation from white (0.0), via yellow (0.33) and orange (0.67), to red (1.0). Similarly, the SSI for sidechain torsions is encoded in the color of the stick representation, and for water sites in the color of the sphere representation.
    For ease of visualization, only side chains corresponding to the twenty largest SSI values are displayed and backbone torsions are represented solely by cartoon.
    Structures are visualized using PyMol.\cite{PyMOL}
    The hydrogen bond featurization for site O3 further reveals how the hydrogen bonding dynamics between the water molecule occupying site O3 and the surrounding protein residues are affected by the change in Asp114 protonation state (C).
    }
    \label{fig:sc-bb-h2o-SSI-muOR}
\end{figure*}

To showcase the usefulness of State-Specific Information (SSI), we investigate the relationship between the protonation state of a central aspartic acid and the $\mu$-opioid receptor ($\mu$OR) ensemble.
The $\mu$OR, a G~protein--coupled receptor (GPCR), is a transmembrane receptor protein that converts extracellular stimuli into intracellular signalling cascades.
The diversity in structure and function among GPCRs underpins complex activation mechanisms that, despite large pharmaceutical interest, remain unresolved.\cite{Katritch2013, Hauser2018, eiger22}
Rotamer changes in residue side chains give rise to larger-scale conformational changes that enable the binding of transducer proteins and trigger downstream signaling (receptor activation).\cite{Hauser2021} These rotamers can be understood as molecular micro-switches, making them an ideal use case for our state-based mutual information approach, SSI.
Many receptors are influenced by environmental pH changes.\cite{Ludwig2003, Rowe2021, Sanderlin2015} Protonation changes of an evolutionarily conserved aspartic acid residue in transmembrane helix 2 (D2.50 in Ballesteros-Weinstein nomenclature\cite{Ballesteros1995}) have been hypothesised to represent a key step in receptor activation.\cite{Ghanouni2000} A simulation study of the $\mathrm{\beta}_2$ adrenergic receptor concluded that it likely becomes protonated upon receptor activation\cite{ranganathan_insights_2014} and a study of the M2 muscarinic receptor linked it to the presence of a proximal Na$^+$ ion.\cite{Vickery2018}
Here, we apply SSI to side-chain and backbone torsions of the antagonist-bound inactive-state murine $\mu$OR (PDB: 4DKL) to investigate the effect of protonating D2.50, i.~e., Asp114 in this particular receptor (Fig.~\ref{fig:sc-bb-h2o-SSI-muOR}).

\emph{Mutual information analysis of side-chain rotamers:}
Multivariate discretization of all backbone and side chain rotamers was performed, identifying Gaussian distributions within each rotamer time-series, similar to the example in Fig.~\ref{fig:multivariate-discretization}. Rotamer states were then identified as Gaussian populations lying within limits defined by the Gaussian intersects for every rotameric angle.  
The SSI values calculated for each residue reveal those parts of the receptor that signal information about the protonation state of Asp114 {by coupled conformational state changes between the rotamer states and the Asp114 protonation state (Fig.~\ref{fig:sc-bb-h2o-SSI-muOR}).
Namely, rotameric state changes in the backbone torsions of transmembrane helix TM6, proximal to Asp114, are coupled to the protonation state changes of Asp114.
Similarly, the backbone rotamer states of intracellular loops ICL2  and ICL3 couple to the protonation state of Asp114.
An outward swing of TM6, enabled by backbone conformational state changes near Phe289 (P-I-F motif), is characteristic of receptor activation,\cite{Katritch2013} and specific conformations of ICL2 and ICL3 are implicated in the binding of signal proteins.\cite{Koehl2018, Mafi2020} 
Furthermore, side chain changes are identified on the P-I-F motif, the NPxxY motif, and the DRY motif, three receptor motifs that are known to undergo distinct rotamer changes in the transition from inactive to active receptor states.\cite{Katritch2013}
The recognition of receptor regions where conformational changes are associated with activation and signaling suggests that the Asp114 protonation state and GPCR activation are intertwined.
This example demonstrates how SSI and its visualization help to pinpoint receptor regions where the features' rotamer states inform about an aspartic acids protonation state.

\emph{Mutual information analysis of water binding sites:}

To further demonstrate how SSI can be used to analyze water molecules, we featurized the ten most well-defined water sites (Fig.~\ref{fig:sc-bb-h2o-SSI-muOR}). 
The locations of the water binding sites were determined using the PENSA water featurizer as the ten sites with the largest probability maxima in the water density grid of the combined ensembles and labelled O1--O10 according to their ranking.
The positioning of all ten sites agrees well with water molecules resolved in experiments\cite{Granier2012, Huang2020} and predicted by the HomolWat server\cite{Mayol2020} for the inactive murine $\mu$OR crystal structure (PDB: 4DKL), confirming the accuracy of PENSA's water site featurization.
Multivariate discretization of all identified water molecules was performed, similar to the example in Fig.~\ref{fig:multivariate-discretization}.
Three water binding sites are within the vicinity of Asp114: O3, O8 and O10.
Using SSI (eq.~\ref{eq:ssi}), we calculated that water binding sites O1--O10 share information with the Asp 114 protonation state on levels between 0.00--0.74 bits.
Water binding site O8, for example, shares 0.67 bits of information in coupled conformational state changes linked to the transition between ensemble $i$ and $j$, i.e., unprotonated to protonated Asp114.
O8 is located beside Phe289 of the P-I-F motif, where we identified that backbone rotamer state changes are coupled to Asp114 protonation.
Surprisingly, the more distant water site O1 shares 0.74 bits of information, 82\% of which is due to an occupancy change. 
Comparing the average ensemble structures reveals an increased packing between TM5 and TM6 in the region about site O1, with the distance between surrounding C$\alpha$s moving over on average 1\AA{} closer in the Asp114-protonated ensemble, suggesting that helix movements on TM6 lead to a collapse of the binding site.
This analysis highlights a concerted behaviour of water binding sites and TM6, whereby state changes to both are indicative of the protonation state of Asp114. It shows how the combined analysis of multiple different features and a comprehensive visualization help to find interrelations within a receptor and discover signaling pathways.

\section{Discussion}

With PENSA, we have implemented an open-source library that provides systematic, easy-to-apply methods that make the otherwise often cumbersome exploration of biomolecular systems faster, more reliable and easier to interpret.
PENSA engages with a range of biomolecular systems via robust featurization implementations. The supported features include interatomic distances and the characteristic torsion angles of amino acids and nucleic acids as well as a novel approach that incorporates water and ion binding sites via their occupancy and, in the case of water, the polarization of the bound water molecule. Combined with dimensionality reduction tools, PENSA can handle biomolecular systems on a wide range of scales and resolutions. 
PENSA includes two comparison measures: JSD and KSS. While the sensitivity of our discrete implementation of the JSD depends on --- and can be adjusted via --- the spacing of the bins, the KSS is a parameter-free metric. Both metrics assess the difference between distributions and in practice often provide similar results, but they differ in their interpretation and typical use case. The JSD tries to answer the question “How different are the distributions?” It works well with large datasets, for which even a fine spacing leaves enough samples in each relevant bin. In contrast, the KSS is more suitable for small datasets, since its usual purpose is hypothesis testing and the comparison of empirical distributions. It tries to answer the question “Are the samples from different distributions?”

In addition, PENSA includes a mutual information measure, State-Specific information (SSI).
With no prior knowledge of the data, PENSA performs an automatic discretization of feature distributions into conformational states, and via SSI, quantifies the information that each features' conformational states signal about the ensemble they are in.
The ideal use case involves features that switch between well-defined states, such as molecular switches, however customizable state definitions allow SSI to operate with many kinds of feature discretization.
SSI can be further extended to three or more features to quantify information flow within a system but is currently only implemented for two. 
Combined with PENSA's convenient visualization tools, these methods allow for a detailed analysis of biomolecular ensembles without the bias of hand-picked metrics, acting as a solid basis for mechanistical interpretations and further, more detailed analysis.

Our example analyses demonstrate the versatility of PENSA on three different biomolecular systems.
By investigating the effects of a small chemical modification on loop opening in an oxidoreductase with JSD, we demonstrate the validity of the method in confirming previously discovered results, while additionally reporting novel, more subtle findings within the same system.
We demonstrate the applicability of PENSA in the optimization of force field parameters via a comparison of the interactions between Calcium and DNA under different force field parameters with KSS.
Finally, we report on a communication channel in the $\mu$-opioid receptor that transmits information between the intracellular signalling site and the protonation state of a distal aspartic acid, shedding further light on the signal transduction mechanism of this mechanistically complicated system.
The major limit to the accuracy of PENSA is the quality of the input ensembles.
For example, insufficiently converged MD simulations can cause false positives when an equally probable transition happens only in one of the conditions. Or they can cause false negatives for overall rare events or slow processes.
If in doubt, validation by other means may be necessary (experiment, independent/longer simulations). Although non-converged simulations can give useful hints, these cases demand a cautious systematic analysis.
Despite these caveats, PENSA has the potential for high-throughput analysis of a large amount of simulations, e.g., those available in GPCRmd\cite{Rodriguez-Espigares2020}, can be used to independently quantify the quality of force fields or generative machine learning models, and can help unravel molecular mechanisms and signaling pathways.


\section{Conclusions}

We present a powerful toolkit to build workflows for the systematic and quantitative analysis of biomolecular systems and their conformational ensembles.
{PENSA's code is open and maintained at \url{https://github.com/drorlab/pensa}.
It provides flexible options to featurize various biomolecular systems, metrics to compare ensembles and to detect interrelations between different regions of a system, and methods to produce intuitive visualizations.
We demonstrate the effectiveness of these methods on three real-world examples from molecular biology, showing how PENSA makes it easier for researchers to analyze large amounts of complex simulation data.


\section{Acknowledgments}

We thank Lukas Stelzl for the trajectory data of the oxidoreductase example as well as Sergio Cruz-Le\'on, Kara Grotz, and Nadine Schwierz for the trajectory data of the DNA-Calcium example. We are  grateful to Alexander Powers, Lukas Stelzl, Nicole Ong, Eleanore Ocana, Emma Andrick, Callum Ives, Bu Tran, and Luca Morlok for beta-testing as well as Masha Karelina, Marc D\"{a}mgen, Patricia Suriana, Sergio Cruz-Le\'on, Michael Ward, and Ramon Guixà-González for helpful discussions. 
M.V. was supported by the EMBO long-term fellowship ALTF 235-2019.
N.J.T. was supported by a BBSRC EASTBIO PhD studentship (grant number BB/M010996/1).
J.M. was supported by the CURIS program of the Stanford University Department of Computer Science.
This work was supported by National Institutes of Health grant R01GM127359 (R.O.D.)
An award of computer time was provided by the INCITE program. This research used resources of the Oak Ridge Leadership Computing Facility, a DOE Office of Science User Facility supported under contract DE-AC05-00OR22725.
Additional computing for this project was performed on the Sherlock cluster and the University of Dundee SLS HPC cluster. We thank Stanford University, the Stanford Research Computing Facility, and the University of Dundee for providing computational resources and support that contributed to these research results.


\section{Author Contributions}
M.V. and N.J.T. conceived of and performed the research and wrote the paper. M.V., N.J.T., S.T.T., and J.M. implemented the PENSA library. U.Z. and R.O.D. supervised the project. M.V. and N.J.T. contributed equally. All authors edited the manuscript and approved of its final version.


\bibliography{pensa}

\end{document}